\documentclass{article}
\usepackage{spconf,amsmath,amssymb,graphicx,booktabs,amsthm}
\usepackage[T1]{fontenc}

\newcommand{\R}{\mathbb{R}}

\newcommand{\ind}{\mathbf{1}}

\usepackage{pifont}
\usepackage{multirow,graphicx}
\usepackage[table]{xcolor}
\definecolor{bestgreen}{RGB}{220,242,220}
 
\definecolor{secondblue}{RGB}{220, 235, 252}

\usepackage{stfloats}
\usepackage{hyperref}
\hypersetup{hidelinks}
\usepackage{xcolor}
\usepackage{soul}
\sethlcolor{gray!25}
\usepackage{cleveref}
\crefname{figure}{Fig.}{Figs.}
\Crefname{figure}{Fig.}{Figs.}
\crefname{equation}{Eq.}{Eqs.}
\Crefname{equation}{Eq.}{Eqs.}

\title{WaveMark: Robust Wavelet-Domain Image Watermarking for Diffusion Models}
\name{Jie Cao$^{1}$ \quad Qi Li$^{2}$ \quad Chaklam Cheong$^{2}$  \quad Yutong Jin$^{1}$ \quad Enze Yu$^{3}$ \quad Zelin Zhang$^{1}$ \quad Jianbing Ni$^{1,}$\sthanks{Corresponding author}}
\address{$^{1}$ Department of Electrical and Computer Engineering,
Queen's University, Kingston, Canada \\
$^{2}$ School of Computing, Queen's University, Kingston, Canada\\
$^{3}$ School of Computer Science, Nanjing University, Jiangsu, China}

\begin{document}
\ninept
\maketitle

\begin{abstract}
Image watermarking enables provenance tracing for photorealistic images generated by diffusion models. However, state-of-the-art (SOTA) removal attacks can suppress embedded signals with little perceptual degradation. This vulnerability limits the effectiveness of watermarks for intellectual property protection and misuse attribution. We propose WaveMark, a novel wavelet-domain watermarking method that embeds messages in a diffusion model's initial noise. To support robust recovery from perturbed carriers, WaveMark distributes keyed replicas of message bits across spatial positions, wavelet orientations, and scales. Each carrier groups coefficients from four latent channels and encodes a randomized bit through magnitude projection, providing an explicit decoding margin. We evaluate WaveMark with SD2.1 on SDP and MS-COCO Captions under 14 attacks. WaveMark achieves 100.0\% bit accuracy (BA) without attacks, while maintaining image quality comparable to the evaluated baselines. Under SOTA MarkNull and MarkNull-A, it retains over 88\% BA, outperforming all baselines.
\end{abstract}


\section{Introduction}
\label{sec:intro}
The growing accessibility of generative AI (GenAI) has made photorealistic image synthesis widely available, expanding opportunities for visual content creation while making image provenance harder to verify. AI-generated images can be circulated as photographic evidence, facilitating impersonation and misinformation and weakening trust in digital media \cite{zhao2024sok}. Watermarking offers a practical means of addressing this challenge by embedding imperceptible, detectable signals that support the identification and attribution of AI-generated images \cite{cao2025secure}. Its importance extends beyond detecting synthetic content to enabling source tracing and accountability across the content distribution process. These needs have driven active research into watermarking methods, including approaches that incorporate identifying signals directly into the image generation process while seeking to preserve generation quality \cite{yang2026t2smark,yang2024gaussian,gunnPRC}. Image watermarking is also seeing wider practical use. Google's SynthID-Image \cite{gowal2025synthid}, for example, is deployed in Gemini, bringing watermarking into a widely used GenAI service to support more transparent and responsible content creation.

In-generation watermarking integrates provenance signals into image synthesis rather than applying them after generation. One line of work fine-tunes model components, such as the VAE decoder, so that each generated image carries a recoverable signature~\cite{fernandez2023stable,xiong2023flexible, kim2024wouaf}. Another line encodes structured patterns or multi-bit messages in the initial noise and recovers them through diffusion inversion~\cite{wen2023tree,yang2024gaussian,gunnPRC}. Both approaches enable source detection and, in some cases, user attribution without a separate post-processing step.

Embedding the watermark during generation, however, does not make it immune to removal. In addition to compression and geometric distortions, regeneration and adversarial attacks can suppress the watermark through generative reconstruction or optimized perturbations while causing little visible degradation~\cite{zhao2024invisible,kassis2025unmarker}. More recent attacks use controllable diffusion-based regeneration~\cite{liu2024image}, manipulate latent representations through surrogate models~\cite{muller2025black,cao2026marknull}, or treat the image as an initial video frame and predict its successor~\cite{qiufuture}. These attacks can remove signals embedded within the generation process itself. Robustness to common distortions is therefore insufficient for provenance verification against a deliberate attacker. A practical watermark must remain detectable under such attacks without reducing image quality.

Advanced removal attacks reduce the reliability of initial-noise watermarks. Regeneration alters the latent signal recovered through inversion, while adversarial perturbations distort the projections used for detection~\cite{zhao2024invisible,an2024waves,kassis2025unmarker, cao2026marknull}. Both effects can move recovered carrier projections toward or across the decision boundary. Reliable bit recovery therefore requires a clear carrier-level margin. \textbf{RQ1: How can we establish a carrier-level decoding margin for reliable bit recovery?} Repeating a bit does not fully solve this problem because replicas exposed to correlated distortions may fail together. This raises \textbf{RQ2: How should bit replicas be distributed to reduce correlated decoding errors?} Increasing the embedding strength widens the separation between bit values, but repeated modifications across carriers can alter the initial-noise energy and reduce image quality. This tradeoff motivates \textbf{RQ3: How can we control carrier energy while preserving the encoded projection?} We address these questions with the following contributions:
\begin{itemize}
    \item We propose WaveMark, a keyed wavelet-domain watermarking framework that distributes randomized bit replicas over disjoint four-channel carriers across spatial positions, wavelet orientations, and scales. 
    \item We introduce projection-magnitude embedding, which maps randomized bits to separated projection magnitudes and scales the orthogonal residual to preserve carrier energy in expectation. This design provides an explicit decoding margin for recovering the embedded watermark after attacks.

    \item   We evaluate WaveMark under 14 attacks on SD2.1. It achieves 100.0\% BA without attacks and retains 88.66\% and 88.06\% BA under MarkNull and MarkNull-A, respectively. We further validate watermark recovery on SD3.5, supporting its applicability across diffusion architectures.
\end{itemize}

\section{Background} 
\textbf{Scenarios.} We consider user attribution in a text-to-image generation service that supports WaveMark. The provider operates the GenAI service. Each compliant user supplies a personal key and a chosen watermark message, then uses the generated images legitimately. During generation, WaveMark uses the supplied key to embed the message into the initial noise. We assume an authenticated association between each user account and its key-message pair. At verification, extracting the watermark with the corresponding key and matching it against the registered message supports attribution to the user who requested the image.

\textbf{Initial-noise-based Watermarking.} WaveMark belongs to initial-noise watermarking \cite{cao2025secure}, which embeds a watermark into the initial latent noise of a fixed latent diffusion model (LDM)~\cite{wen2023tree,yang2024gaussian,gunnPRC,yang2026t2smark}. Given a secret user key $K$, a binary message $m$, and a text prompt $y$, image generation is formulated as:
\begin{equation}
z_T \sim \mathcal{N}(\mathbf{0},\mathbf{I}), \;
z_T^w = \mathsf{Embed}(z_T;K,m), \;
I_w = \Theta(z_T^w,y),
\end{equation}
where $\mathsf{Embed}$ constructs the watermarked initial noise and $\Theta$ denotes the generative model. The embedding should maintain generation quality and consistency with the text prompt $y$.

To decode the message in a query image $I$, we first estimate its initial noise through approximate diffusion inversion, such as DDIM~\cite{song2020denoising} or DPM-Solver~\cite{lu2025dpm}, and then extract the message using the key $K$. With DDIM inversion, this process is expressed as:
\begin{equation}
    \hat{z}_T = \mathsf{DDIM}_{0\rightarrow T}(I;\Theta),
    \qquad
    \hat{m} = \mathsf{Extract}(\hat{z}_T,K).
\end{equation}
The extracted message $\hat{m}$ is compared with the target message $m$ using bit accuracy (BA), which measures the ratio of correctly recovered bits:
\begin{equation}
    \mathrm{BA}(\hat{m},m)
    = 1-\frac{d_H(\hat{m},m)}{n},
    \label{eq:ba}
\end{equation}
where $d_H$ denotes the Hamming distance and $n$ is the message length. We accept the image as carrying the target watermark if $\mathrm{BA}(\hat{m},m)\geq\tau$. To choose the verification threshold $\tau$, we follow~\cite{fernandez2023stable} and specify a target false-positive rate (FPR) $\varepsilon$, requiring
\begin{equation}
    \mathrm{FPR}(\tau)
    = \Pr\!\left(
        \mathrm{BA}(\hat{m},m)\geq\tau \mid H_0
      \right)
    \leq \varepsilon,
    \label{eq:fpr}
\end{equation}
where $H_0$ denotes the absence of the target watermark. Under $H_0$, we assume that the extracted bits are independent and uniformly distributed, so the number of matches with the $n$-bit reference message follows $\mathsf{Binomial}(n,0.5)$. We select $\tau$ from the upper tail of this distribution to satisfy the FPR constraint.


\section{Methodology}
\label{sec:embed}
\subsection{Overview}
\begin{figure}[t]
    \centering
    \includegraphics[width=1\linewidth]{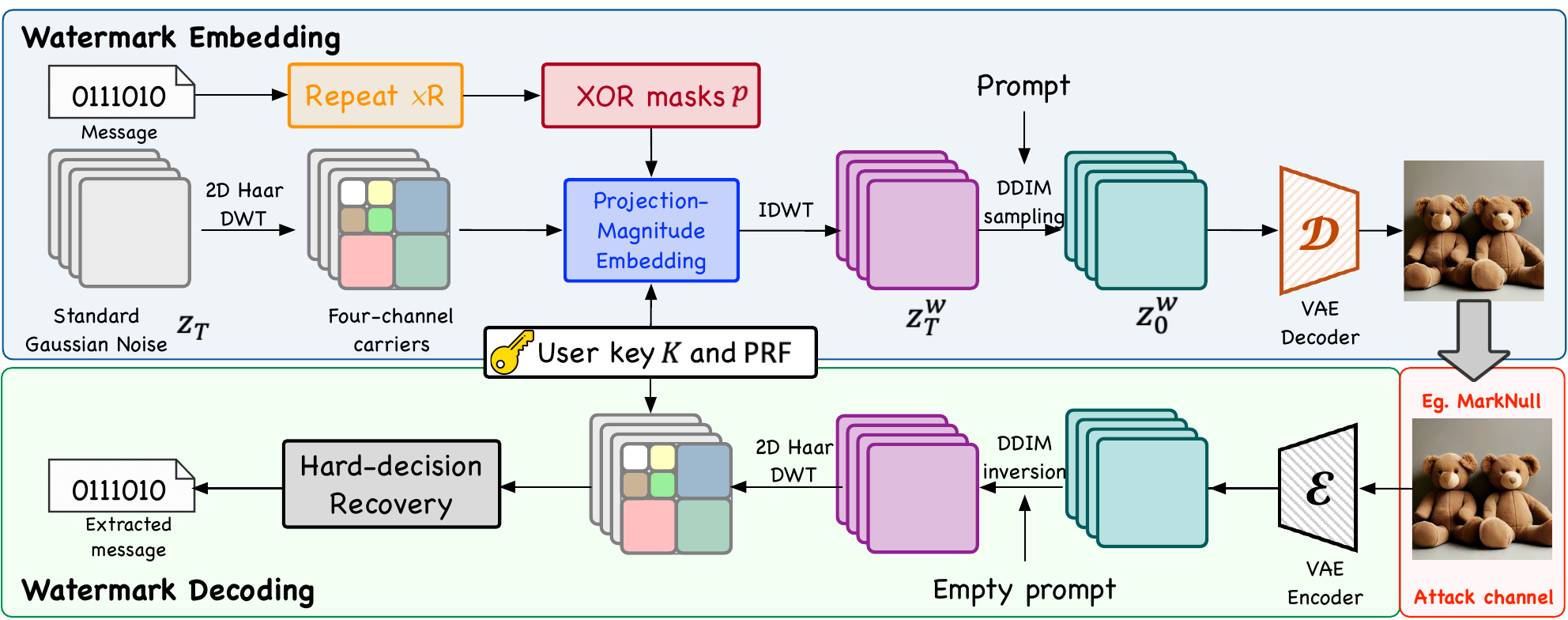}
    \caption{Overview of the WaveMark framework.}
    \label{fig:overview}
\end{figure}

Fig.~\ref{fig:overview} illustrates the embedding and decoding pipelines of WaveMark. We first repeat each message bit $R$ times and apply XOR randomization derived from the user key $K$. A two-level Haar Discrete Wavelet Transform (DWT) decomposes the initial noise $z_T\sim\mathcal{N}(0,\mathbf{I})$ into wavelet subbands, where a keyed Pseudorandom Function ($\mathsf{PRF}$) assigns the replicas to disjoint four-channel carriers across spatial positions, orientations, and scales. For each carrier, projection-magnitude embedding sets the projection magnitude along a keyed direction to zero or $\alpha$, according to the randomized bit. We scale the orthogonal residual to preserve carrier energy in expectation while leaving the encoded projection unchanged. Inverse DWT then reconstructs the watermarked noise $z_T^w$, which undergoes prompt-conditioned DDIM sampling and VAE decoding to generate the watermarked image. At decoding, the VAE encoder maps an image to its latent representation, and DDIM inversion with an empty prompt estimates the initial noise. We apply the same Haar DWT and use $K$ to locate the carriers and reproduce their projection directions. The decoder compares each projection magnitude with $\alpha/2$, reverses the XOR randomization, and combines the replica decisions through hard majority voting to recover the message.

\subsection{Message Bits and Keyed Wavelet Carriers}
\label{sec:carriers}
We apply a two-level orthonormal Haar DWT independently to each channel of $z_T\in\mathbb{R}^{4\times64\times64}$. The selected pools are $\mathrm{HL}_1$, $\mathrm{LH}_1$, $\mathrm{HL}_2$, and $\mathrm{LH}_2$. At one selected spatial position in a pool, the $d=4$ coefficients across channels form a carrier $c_{j,r}\in\R^4$ for watermark message $m=(m_j)_{j=1}^{n}\in\{0,1\}^{n}$ and replica index $r\in\{0,\ldots,R-1\}$. Each bit has $R$ carriers. Selecting positions without replacement within each pool gives $M=nR$ disjoint carriers containing $4M$ scalar coefficients. The coarsest approximation band, diagonal detail bands, and all unselected coefficients remain unchanged.

\begin{equation}
\mathsf{DWT}^{(2)}(z_T)=\{\mathrm{LL}_2,\{\mathrm{HL}_l,\mathrm{LH}_l,\mathrm{HH}_l\}_{l=1}^{2}\}.
\label{eq:dwt-decomposition}
\end{equation}

We distribute replicas across the $\mathrm{HL}$ and $\mathrm{LH}$ subbands at both wavelet levels. For each replica $(j,r)$, we generate a binary mask $p_{j,r}\in\{0,1\}$ and a unit direction vector $a_{j,r}\in\{-1/2,+1/2\}^{4}$ using domain-separated outputs of $\mathsf{PRF}_K$, implemented with HMAC-SHA256 and keyed by the user key $K$. The mask determines the bit to be embedded, $b_{j,r}=m_j\oplus p_{j,r}$, where $\oplus$ denotes XOR: a zero mask preserves the message bit, while a one mask flips it. The direction vector $a_{j,r}$ provides the signed weights used to combine the carrier's four channel coefficients into a scalar projection, whose magnitude is modified to embed $b_{j,r}$.

Specifically, let $\mathcal{S}_{j,r}$ denote the subband assigned to replica $r$ of message bit $m_j$. Within this subband, we assign the carrier a spatial position using the following allocation rule based on $\mathsf{PRF}_K$:
\vspace{-0.5em}
\begin{equation} 
(\rho_{j,r},\eta_{j,r})=\mathsf{PRF}^{\mathrm{pos}}_K(\mathcal{S}_{j,r},j,r),\;\mathcal{S}_{j,r}\in\{\mathrm{HL}_l,\mathrm{LH}_l\}_{l=1}^{2}. \label{eq:carrier-position} \vspace{-0.5em}\end{equation}

Here, $\mathsf{PRF}^{\mathrm{pos}}_K$ denotes the keyed position allocation procedure, which ranks candidate positions using $\mathsf{PRF}_K$ outputs and assigns them without replacement within each subband. The same key and configuration reproduce this allocation at extraction. For example, when $\mathcal{S}_{j,r}=\mathrm{HL}_1$, the subband has shape $\mathrm{HL}_1\in\mathbb{R}^{4\times32\times32}$ and contains four $32\times32$ coefficient maps, one per latent channel. We collect the coefficient at position $(\rho_{j,r},\eta_{j,r})$ from each map to form the carrier:
\vspace{-0.5em}
\begin{equation} 
c_{j,r}= \bigl(\mathrm{HL}_{1}^{(i)}[\rho_{j,r},\eta_{j,r}]\bigr)_{i=1}^{4}\in\mathbb{R}^{4},
\label{eq:carrier-example}\vspace{-0.5em} \end{equation} 
where \(i\) indexes latent channels, and \((\rho_{j,r},\eta_{j,r})\) is the keyed position where \(c_{j,r}\) groups four cross-channel coefficients to embed \(b_{j,r}\). 
\hl{\textbf{To RQ2:} The keyed assignment places replicas in disjoint
spatial carriers from the \mbox{$\mathrm{HL}$} and \mbox{$\mathrm{LH}$} subbands at both wavelet scales. This distribution reduces correlated decoding errors.}

\subsection{Projection-Magnitude Embedding} \label{sec:embedding}
Building on the projection-modulation principle of spread-transform dither modulation (STDM)~\cite{chen1999achievable}, we embed each randomized bit \(b_{j,r}\) by jointly modifying the four cross-channel coefficients of its selected carrier \(c_{j,r}\). Specifically, we set the projection magnitude along \(a_{j,r}\) to \(0\) for \(b_{j,r}=0\) and to a fixed amplitude \(\alpha\) for \(b_{j,r}=1\), where \(\alpha>0\) controls the embedding strength. For clarity, we omit the indices \((j,r)\) and write \(c=c_{j,r}\), \(a=a_{j,r}\), and \(b=b_{j,r}\). We decompose the original carrier \(c\) as follows:

\vspace{-0.5em}
\begin{equation} 
s=a^\top c=\sum_{i=1}^{4}a_i c_i, \qquad v=c-sa. \label{eq:decomp} 
\end{equation} \vspace{-0.5em}

Here, $s$ is the scalar projection of $c$ along the unit direction $a$, and $v$ contains the remaining component. Since $\|a\|_2=1$, we have $a^\top v=0$ and $c=sa+v$. This decomposition allows us to change the projection that carries the bit while separately controlling the energy of the orthogonal component. We choose the target projection according to the randomized bit: 

\begin{equation} q=\begin{cases} 0, & b=0,\\ \alpha\,\operatorname{sgn}_{+}(s), & b=1, \end{cases} \label{eq:target} 
\end{equation} 
where $\operatorname{sgn}_{+}(s)=-1$ for $s<0$ and $+1$ otherwise. For $b=1$, this rule selects the closer of $-\alpha$ and $+\alpha$, minimizing $|q-s|$; when $s=0$, we choose $+\alpha$. Both nonzero targets represent randomized bit one, so their sign carries no message information. We then replace the original parallel component $sa$ with $qa$ and scale the orthogonal component by a fixed gain $g>0$: 

\begin{equation}c^w=qa+gv=qa+g(c-sa).
\label{eq:amplitude_embed} \end{equation}

The gain \(g\) compensates for the energy introduced by the target projection. Since \(qa\) and \(gv\) are orthogonal,

\begin{equation}
\|c^w\|_2^2=q^2+g^2\|v\|_2^2.
\end{equation}

For a \(4\)-dimensional standard Gaussian carrier, \(\mathbb E[\|c\|_2^2]=4\) and \(\mathbb E[\|v\|_2^2]=3\). The keyed XOR masks make the randomized bits balanced in expectation, giving \(\mathbb E[q^2]=\alpha^2/2\). To preserve the expected carrier energy, we require:

\begin{equation}
\frac{\alpha^2}{2}+3g^2=4
\label{eq:g_cal}
\end{equation}

We define the residual-energy retention ratio as \(\delta=g^2\). A smaller \(\delta\) retains less of the original residual energy but permits a larger \(\alpha\), providing a wider watermark decision margin under the energy constraint above. In WaveMark, we set \(\delta=1/2\), which gives \(\alpha=\sqrt{5}\) and \(g=1/\sqrt{2}\). \hl{\textbf{To RQ3:} The gain \mbox{$g$} controls residual-energy
retention without changing the encoded projection, while the energy
constraint preserves the expected carrier energy.}


Finally, we write each updated carrier $c_{j,r}^{w}$ back to its selected position in the assigned subband $\mathcal{S}_{j,r}$:
\vspace{-0.5em}
\begin{equation}
\bigl(\mathcal{S}_{j,r}^{w}\bigr)^{(i)}
[\rho_{j,r},\eta_{j,r}]
\leftarrow
\bigl(c_{j,r}^{w}\bigr)_i,
\quad i=1,\ldots,4.
\label{eq:carrier-writeback}
\end{equation}
where \(\mathcal{S}_{j,r}^{w}\) is the updated subband. We modify only the selected carriers and apply the inverse Haar DWT to obtain \(z_T^{w}\) for diffusion sampling and text-to-image generation.

\subsection{Watermark Decoding}
\label{sec:extract}

For a query image $I$, we obtain $z_0$ from the VAE encoder $\mathcal{E}$ using the posterior mode and the model's latent scaling factor. We then run DDIM inversion with an empty prompt to estimate $\hat z_T=\mathsf{DDIM}_{0\rightarrow T}(z_0)$. After applying the same Haar DWT, we use $\mathsf{PRF}_K$ and $\mathsf{PRF}^{\mathrm{pos}}_K$ to reconstruct the carrier positions and projection directions. For each recovered carrier $\hat c_{j,r}$, we compute $\hat s_{j,r}=a_{j,r}^{\top}\hat c_{j,r}$ and compare its magnitude $|\hat s_{j,r}|$ with $\alpha/2$, the midpoint between the two target magnitudes:
\begin{equation}
 \hat b_{j,r}=\ind\bigl[|\hat s_{j,r}|>\alpha/2\bigr],\;h_{j,r}=\hat b_{j,r}\oplus p_{j,r},
 \label{eq:hard}
\end{equation}
where $\ind[\cdot]$ equals $1$ when the condition holds and $0$ otherwise. We first estimate the randomized bit, then remove the randomization to obtain the replica's vote $h_{j,r}$ for message bit $m_j$. We count the votes for 1 for each message bit and apply the following fixed voting rule:
\begin{equation}
 N_j=\sum_{r=0}^{R-1}h_{j,r},
 \qquad \hat m_j=\ind\bigl[N_j\geq R/2\bigr].
 \label{eq:majority}
\end{equation}

A message bit is decoded as $1$ when at least $\lceil R/2\rceil$ replicas vote for $1$. For even $R$, ties are resolved in favor of $1$. We evaluate the extracted message $\hat m$ against the original message $m\in\{0,1\}^{n}$ based on \cref{eq:ba}. \hl{\textbf{To RQ1:} The target magnitudes 0 and \mbox{$\alpha$} place
the decision threshold at \mbox{$\alpha/2$}, so both targets have a
margin of \mbox{$\alpha/2$}.}


\begin{table*}[t]
\centering
\caption{Watermark performance and image quality on SDP and MS-COCO Captions, with 100 images per method per dataset.}
\label{tab:watermark_comparison}
\setlength{\tabcolsep}{4pt}
\renewcommand{\arraystretch}{1}
\resizebox{0.85\textwidth}{!}{%
\begin{tabular}{l*{8}{c}}
\toprule
\multirow{2}{*}{Method}
& \multicolumn{4}{c}{SDP}
& \multicolumn{4}{c}{MS-COCO Captions} \\
\cmidrule(lr){2-5}\cmidrule(lr){6-9}
& BA (\%) $\uparrow$
& CLIP-Score $\uparrow$
& CLIP-IQA $\uparrow$
& FID $\downarrow$
& BA (\%) $\uparrow$
& CLIP-Score $\uparrow$
& CLIP-IQA $\uparrow$
& FID $\downarrow$ \\
\midrule
SSig$_{\text{[ICCV'23]}}$
& 99.85 & 27.689 & 0.8452 & 155.680
& 99.69 & 26.021 & 0.8937 & 153.196 \\
SleeperMark$_{\text{[CVPR'25]}}$
& 99.92 & 28.229 & 0.8796 & 169.536
& 99.50 & 26.381 & 0.8421 & 161.515 \\
VINE$_{\text{[ICLR'25]}}$
& 99.84 & 27.066 & 0.9143 & 153.068
& 99.82 & 25.760 & 0.9422 & 151.344 \\
GS$_{\text{[CVPR'24]}}$
& 100.0 & 27.701 & 0.9064 & 151.392
& 100.0 & 25.799 & 0.9479 & 141.264 \\
PRC$_{\text{[ICLR'25]}}$
& 100.0 & 28.047 & 0.9104 & 156.512
& 100.0 & 26.034 & 0.9405 & 147.439 \\
T2SMark$_{\text{[NeurIPS'25]}}$
& 100.0 & 27.946 & 0.9033 & 151.837
& 100.0 & 25.832 & 0.9392 & 148.548 \\
\midrule
\textbf{WaveMark}
& 100.0 & 27.648 & 0.9244 & 153.926
& 100.0 & 25.680 & 0.9476 & 151.412 \\
\bottomrule
\end{tabular}%
}
\end{table*}

\begin{table*}[t]
\centering
\caption{BA (\%) of watermarking methods under various attacks.}
\label{tab:robustness_ba}
\renewcommand{\arraystretch}{1}
\setlength{\tabcolsep}{4pt}
\resizebox{\textwidth}{!}{%
\begin{tabular}{l*{14}{c}}
\toprule
\multirow{2}{*}{Watermarking}
& \multicolumn{5}{c}{Common image distortions}
& \multicolumn{4}{c}{Regeneration-based attacks}
& \multicolumn{5}{c}{Adversarial attacks} \\
\cmidrule(lr){2-6}
\cmidrule(lr){7-10}
\cmidrule(lr){11-15}
& Noise & Blur &Crop & Contrast & JPEG
& DA & VA & CtrlRegen+ & NFPA
& UnMarker & Imprint & NPA & MarkNull & MarkNull-A \\
\midrule
SSig & 95.08 & 84.33 &99.16 &98.04 & 85.13 & 48.50 & 62.79 & 46.63 & 48.50 &92.18 & 45.83 & 47.25 & 49.33 & 45.63 \\
SleeperMark & 98.92 & 99.88 &95.91 &99.70 & 99.83 & 99.46 & 95.75 & 88.04 & 82.79 &99.52 & 83.23 & 90.17 & 41.63 & 41.38 \\
VINE & 99.30 & 99.64 &52.42 &98.08 & 99.82 & 90.04 & 95.16 & 71.06 & 50.88 &97.54  & 71.70 & 80.26 & 61.36 & 57.16 \\
GS & 99.59 & 100.0 &99.72 &99.99 & 99.97 & 99.68 & 98.76 & 93.35 & 51.16 &99.83 & 28.16 & 49.27 & 36.49 & 74.74 \\
PRC & 70.64 & 82.18 &66.94 &80.35 & 78.83 & 69.38 & 66.51 & 59.42 & 50.13 &91.46 & 46.45 & 49.89 & 47.41 & 54.48 \\
T2SMark & 99.53 & 100.0 &99.96 &99.97 & 99.92 & 99.90 & 98.14 & 94.62 & 49.83 &98.17 & 52.18 & 63.92 & 51.82 & 51.05 \\
\midrule
\textbf{WaveMark} & 98.84 & 100.0 &99.75 &99.81 &99.94 & 98.25 & 96.72 & 79.47& 54.18 & 98.65& 84.72 & 92.47 & 88.66 & 88.06 \\
\bottomrule
\end{tabular}%
}
\end{table*}

\section{Experiments}
\subsection{Experimental Setup}
\textbf{WaveMark Setup.} We generate $512\times512$ images using Stable Diffusion 2.1-base (SD2.1)~\cite{stabilityai_sd21base_2022} with 50 DDIM sampling steps and a guidance scale of 7.5. WaveMark embeds a fixed 64-bit message using a 256-bit key. We assign \(12\), \(12\), \(4\), and \(3\) replicas per bit to \(\mathrm{HL}_1\), \(\mathrm{LH}_1\), \(\mathrm{HL}_2\), and \(\mathrm{LH}_2\) (i.e., $R=31$), yielding \(1{,}984\) disjoint carriers for the 64-bit message. For decoding, we run 50 DDIM inversion steps with an empty prompt and a guidance scale of 1.0. Using \cref{eq:fpr}, we set \(\epsilon=10^{-3}\) and use \(\tau=46/64=71.875\%\), corresponding to an FPR of \(3.09\times10^{-4}\) under \(H_0\).

\textbf{Models and datasets.}
We use SD2.1 as the default backbone for image generation and for watermarking methods that operate during diffusion generation. Prompts are randomly sampled from the Stable Diffusion Prompts (SDP) dataset~\cite{gustavosta_stable_diffusion_prompts} and MS-COCO Captions dataset \cite{chen2015microsoft}. For each in-generation method, we generate 100 watermarked images at a resolution of $512\times512$ using SD2.1. All diffusion generation runs use 50 DDIM denoising steps. 

\textbf{Watermarking Baselines.} We compare our method with six baselines for watermarking AI-generated images, including Stable Signature (SSig)~\cite{fernandez2023stable}, SleeperMark~\cite{wang2024sleepermark}, VINE~\cite{lu2024robust}, Gaussian Shading (GS)~\cite{yang2024gaussian}, PRC~\cite{gunnPRC}, and T2SMark~\cite{yang2026t2smark}.

\textbf{Attack baselines.}
We evaluate WaveMark against five common image distortions: Gaussian noise, Gaussian blur, 50\% random cropping, contrast adjustment, and JPEG compression. We also consider regeneration-based removal attacks, including DA~\cite{zhao2024invisible}, VA~\cite{an2024waves}, CtrlRegen+~\cite{liu2024image}, and NFPA~\cite{qiufuture}, as well as adversarial attacks, including UnMarker~\cite{kassis2025unmarker}, Imprint~\cite{muller2025black}, and NPA~\cite{goren2025noiseprints}. We additionally evaluate robustness against two recent SOTA removal attacks: MarkNull and MarkNull-A~\cite{cao2026marknull}. We follow the attack configurations reported in the original papers and apply each attack to 100 generated watermarked images per watermarking method.

\textbf{Evaluation Metrics.} We evaluate watermark performance using BA. We assess image generation quality using CLIP-Score~\cite{hessel2021clipscore}, CLIP-IQA~\cite{wang2023exploring}, and FID~\cite{heusel2017gans}.

\subsection{Experimental Results}

\begin{figure}[h]
    \centering
    \includegraphics[width=\linewidth]{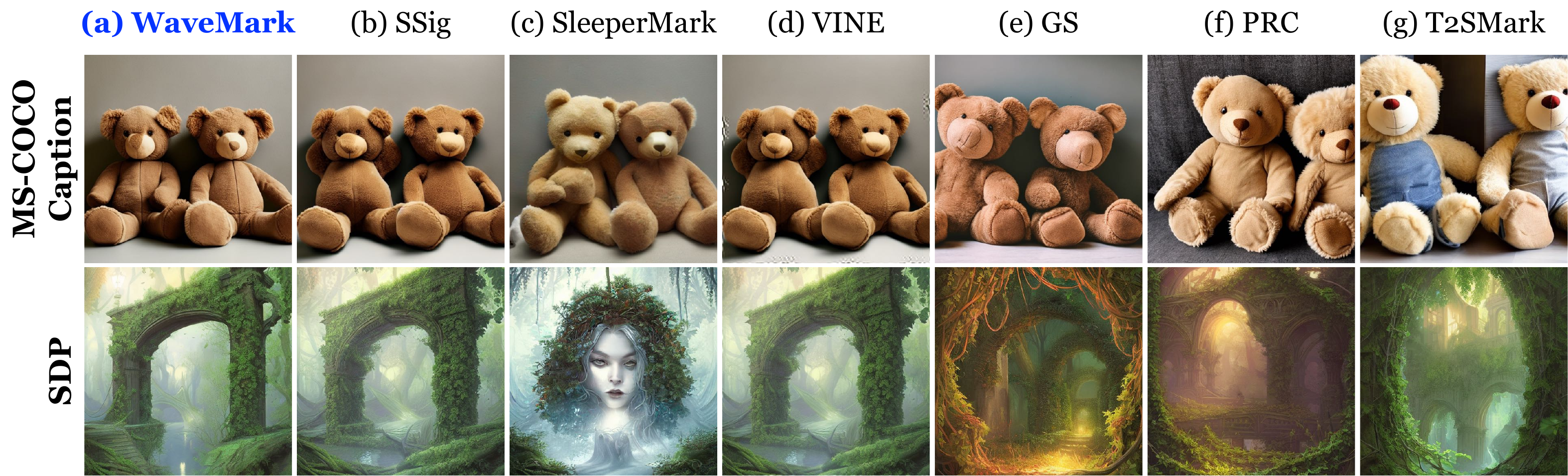}
    \caption{Images generated by different watermarking methods using SD2.1 on two prompt datasets.}
    \label{fig:sample}
    \vspace{-1em}
\end{figure}

\textbf{Watermark performance and image quality.} As shown in Table~\ref{tab:watermark_comparison}, WaveMark reaches 100.0\% BA on both datasets. WaveMark has the highest CLIP-IQA on SDP and ranks second on MS-COCO Captions, which suggests that embedding the watermark has little effect on perceived image quality. Its CLIP-Score decreases from 27.648 on SDP to 25.680 on MS-COCO Captions. The baselines show a similar decrease, so this change appears to be associated with the dataset rather than with WaveMark alone. The corresponding FID values are 153.926 and 151.412. WaveMark does not have the best value for every metric, but it retains competitive image quality while achieving full BA. Figure~\ref{fig:sample} provides a visual comparison of images generated by different watermarking methods using SD2.1. The WaveMark samples preserve both semantic content and fine visual details, with no obvious watermark-induced artifacts, demonstrating promising image quality.

\textbf{Robustness Evaluation.} Table~\ref{tab:robustness_ba} compares WaveMark with six baselines under 14 attacks. WaveMark achieves 98.84\% to 100.0\% BA under common distortions, including 99.75\% under 50\% random cropping with black padding and 99.81\% under contrast adjustment. Under regeneration attacks, BA reaches 98.25\% for DA and 96.72\% for VA, but decreases to 79.47\% for CtrlRegen+ and 54.18\% for NFPA. Under the SOTA MarkNull and MarkNull-A attacks, WaveMark achieves the highest BA of 88.66\% and 88.06\%, respectively, improving on the strongest baselines by 27.30\% and 13.32\%. NFPA treats the watermarked image as an initial video frame and predicts the next frame to remove the watermark. Although this suppresses the watermark, the resulting images differ substantially from the original inputs~\cite{cao2026marknull,qiufuture}.Except for NFPA, WaveMark's BA remains above the 71.875\% detection threshold under all evaluated attacks. These results demonstrate strong robustness to common distortions and adversarial removal attacks, providing a practical foundation for reliable watermark detection and user attribution.

\textbf{Cross-architecture Compatibility.} To assess compatibility across diffusion architectures, we evaluate WaveMark on Stable Diffusion 3.5-Medium (SD3.5) \cite{esser2024scaling}, extending our experiments from the U-Net-based SD2.1 to the DiT architecture. WaveMark achieves 99.81\% and 100.0\% BA on SDP and MS-COCO Captions. The corresponding CLIP scores are 28.963 and 26.513, CLIP-IQA scores are 0.937 and 0.922, and FID values are 81.079 and 205.881. These results demonstrate reliable watermark recovery and detection on SD3.5, supporting WaveMark’s compatibility with both U-Net and DiT architectures.

\section{Conclusion}
We have proposed WaveMark, a robust wavelet-domain watermarking method that embeds messages in the initial noise of LDMs. It distributes keyed message replicas across wavelet subbands and encodes each replica in a carrier's projection magnitude to improve the robustness against advanced watermark removal attacks. On SD2.1, WaveMark achieves 100.0\% BA without attacks and retains 88\% BA under MarkNull and MarkNull-A, outperforming all baselines. Experiments on SD3.5 also support its applicability to a DiT architecture for watermark embedding. Future work will extend WaveMark toward publicly verifiable watermarking, enabling third-party verification without disclosing users’ secret keys.
\bibliographystyle{IEEEbib}
\bibliography{refs}
\end{document}